\documentclass[12pt]{article}

\usepackage{graphicx}
\usepackage{amsmath,amssymb}
\usepackage{bm}
\usepackage{graphicx, color}
\usepackage{wrapfig}
\begin{document}
\title{
\begin{flushright}
\ \\*[-80pt] 
\begin{minipage}{0.2\linewidth}
\normalsize
%arXiv:YYMM.NNNN \\
%KUNS-xxxx \\*[50pt]
\end{minipage}
\end{flushright}
{\Large \bf 
$S_4$ Flavor Model of Quarks and Leptons
\\*[10pt]}}

\author{
\centerline{
Hajime~Ishimori$^{1}$,
~Yusuke~Shimizu$^{1,}$\footnote{talked at the 16th Yukawa International Seminar, Particle Physics beyond the Standard Model, Yukawa Institute for Theoretical Physics, in Kyoto, Japan, January 26 - March 25, 2009}, } \\ 
\centerline{
Morimitsu~Tanimoto$^{2}$}
\\*[15pt]
\centerline{
\begin{minipage}{\linewidth}
\begin{center}
$^1${\it \normalsize
Graduate~School~of~Science~and~Technology,~Niigata~University, \\ 
Niigata~950-2181,~Japan } \\
$^2${\it \normalsize
Department of Physics, Niigata University,~Niigata 950-2181, Japan } 
\end{center}
\end{minipage}}
\\*[25pt]}

\date{
\centerline{\small \bf Abstract}
\begin{minipage}{0.9\linewidth}
\medskip 
\medskip 
\small
 We  present  an $S_4$ flavor model to unify
  quarks and leptons in the framework of the $SU(5)$ GUT.
 Three generations of $\overline 5$-plets in $SU(5)$ are assigned $3_1$
of $S_4$ while the  first and the second generations of 
$10$-plets  in  $SU(5)$  are assigned to be  $2$ of $S_4$,
and the third generation of $10$-plet is to be  $1_1$ of $S_4$.
Right-handed neutrinos are also assigned  
 $2$ for the first and second generations
and $1_1$ for  the third generation, respectively.
 Taking vacuum alignments of  relevant gauge singlet scalars,
 we predict  the  quark  mixing  as well as the tri-bimaximal
mixing of neutrino flavors. Especially, the Cabbibo angle is
predicted to be $15^{\circ}$ in the limit of  the vacuum alignment.
We can improve the model to predict  observed CKM mixing angles.
\end{minipage}
}

\maketitle
\vspace{-0.8cm}
\section{Introduction}

Neutrino experimental  data provide  an important clue for elucidating the 
origin of the observed hierarchies in mass matrices for quarks and leptons.
Recent experiments of the neutrino oscillation 
go into a  new  phase  of precise  determination of
 mixing angles and mass squared  differences  \cite{Threeflavors}, 
which indicate the tri-bimaximal mixing  for three flavors 
 in the lepton sector  \cite{HPS}. 
These large  mixing angles are %completely  
 different from the quark mixing ones.
Therefore, it is %very 
important
to find a natural model that leads to these mixing patterns
 of quarks and leptons with good accuracy.

 Flavor symmetry is expected to explain  the mass spectrum  and
the mixing matrix of quarks and leptons. 
In particular,  some predictive models with  non-Abelian  discrete flavor
symmetries have been explored by many authors.
Among them,
 the tri-bimaximal mixing of leptons has been   understood 
based on the non-Abelian finite group 
$A_4$, {\cite{A4-Ma-and-Rajasekaran}-\cite{A4-Hirsch3} 
because   these  symmetries provide the definite meaning of generations
 and connects different generations.
On the other hand, much  attention has been devoted to the question $SU(5)$
whether these  models  can be  extended to
describe the observed pattern of quark masses and mixing angles,
and whether these can be made compatible with the $SU(5)$ or $SO(10)$ 
grand unified theory (GUT).

Recently,   group-theoretical arguments indicate that
the discrete symmetry   $S_4$ is the minimal flavor
symmetry compatible with the tri-bimaximal neutrino mixing \cite{Lam}.
Actually, the exact  tri-bimaximal neutrino mixing is realized
 in the $S_4$ flavor model \cite{S4-Morisi}.
Thus, the $S_4$ flavor model is attractive for the lepton sector.
\cite{S4-Ma}-\cite{S4-Koide}
Although an  attempt to unify quark and lepton sector was presented
towards a grand unified theory of flavor \cite{S4-Hagedorn}, 
  mixing angles are not predicted clearly.
In this work, we present an $S_4$ flavor model to unify
 the quarks and leptons in the framework of the $SU(5)$ GUT.

\section{Prototype of  $S_4$ flavor model in $SU(5)$ GUT}

We present the prototype of  the  $S_4$ flavor model 
in the $SU(5)$ GUT to understand the essence of our model clearly.
We consider the supersymmetric GUT based on $SU(5)$
The flavor symmetry of quarks and leptons is the discrete group $S_4$
 in our model.
 The group $S_4$ is the permutation group of four distinct objects \cite{Lomont, Califano}.
It is isomorphic to the symmetry group of regular octahedron.
It has 24 distict elements and has five irreducible representations 
$1_1,~1_2,~2,~3_1,$ and $3_2$.
%The group $S_4$ has irreducible representations $1_1,~1_2,~2,~3_1,$ and $3_2$. The multiplication rules of  $S_4$ are summarized in appendix.

\begin{table}[h]
\begin{tabular}{|c|ccccc||cc|}
\hline
&$(T_1,T_2)$ & $T_3$ & $( F_1, F_2, F_3)$ & $(N_e^c,N_\mu ^c)$ & $N_\tau ^c$ & $H_5$ &$H_{\bar 5} $ \\ \hline
$SU(5)$ & $10$ & $10$ & $\bar 5$ & $1$ & $1$ & $5$ & $\bar 5$ \\
$S_4$ & $2$ & $1_1$ & $3_1$ & $2$ & $1_1$ & $1_1$ & $1_1$ \\
$Z_4$ & $\omega ^3$ & $\omega ^2$ & $\omega $ & $1$ & $1$ & $1$ & $1$ \\
\hline
\end{tabular}
\end{table}
\vspace{-0.8cm}
\begin{table}[h]
\begin{tabular}{|c|cccccc|}
\hline
& $\chi _1$&  $(\chi _2,\chi _3)$ &  $(\chi _4,\chi _5)$&$(\chi _6,\chi _7,\chi _8)$  & $(\chi _9,\chi _{10},\chi _{11})$ & $(\chi _{12},\chi _{13},\chi _{14})$ \\ \hline
$SU(5)$ & $1$ & $1$ & $1$ & $1$ & $1$ & $1$ \\
$S_4$ & $1_1$ & $2$ & $2$ & $3_1$ & $3_1$ & $3_1$ \\
$Z_4$ & $\omega^2$ & $\omega^2$ & $1$ & $\omega ^3$ & $1$ & $\omega $ \\
\hline
\end{tabular}
\caption{Assignments of $SU(5)$, $S_4$, and $Z_4$ representations, 
where  the phase factor $\omega$ is $i$.}
\end{table}

Let us present the model of the quark and lepton  flavor 
with the $S_4$ group in $SU(5)$ GUT. 
In $SU(5)$, matter fields are unified into a $10(q_1,u^c,e^c)_L$ 
and a $\bar 5(d^c,l_e)_L$ dimensional representations. 
Three generations of $\bar 5$, which are denoted by $F_i$,
 are assigned by $3_1$ of $S_4$. 
On the other hand, the third generation of the $10$-dimensional 
representation is assigned by $1_1$ of $S_4$, 
so that the top quark Yukawa coupling is allowed in tree level. 
While, the first and the second generations are assigned $2$ of $S_4$. 
These $10$-dimensional representations are denoted by 
$T_3$ and $(T_1,T_2)$, respectively.
Right-handed neutrinos, which are $SU(5)$ gauge singlets,
are also assigned  $1_1$ and  $2$ for $N^c_\tau$ and  
$(N^c_e,N^c_\mu)$, respectively.
 
We introduce new scalars 
$\chi_i$ in addition to the $5$-dimensional 
and $\bar 5$-dimensional Higgs of the $SU(5)$, $H_5$ and $H_{\bar 5} $
 which are  assigned $1_1$ of $S_4$. 
These new scalars are supposed to be $SU(5)$ gauge singlets. 
The  $\chi_1$ scalar is assigned $1_1$, $(\chi _2,\chi _3)$, 
and $(\chi _4,\chi _5)$ are $2$, $(\chi _6,\chi _7,\chi _8),~(\chi _9,\chi _{10},\chi _{11})$, and $(\chi _{12},\chi _{13},\chi _{14})$ 
are $3_1$ of the $S_4$ representations, respectively. 
The  $\chi_1$ and $(\chi _2,\chi _3)$  scalars
are coupled with the  up-type  quark sector, $(\chi _4,\chi _5)$ are  
coupled with the right-handed Majorana neutrino sector, 
$(\chi _6,\chi _7,\chi _8)$ are  coupled with the Dirac neutrino sector, 
$(\chi _9,\chi _{10},\chi _{11})$ and $(\chi _{12},\chi _{13},\chi _{14})$ 
are coupled with the charged lepton and down-type quark sector, respectively. 
We also add $Z_4$ symmetry in order to obtain   relevant couplings.
The particle assignments of $SU(5)$, $S_4$, and $Z_4$ are summarized Table 1.

 We can now write down  the superpotential at the leading order 
 in terms of the cut off scale $\Lambda$, 
which is taken to be the Planck scale. The $SU(5)$ invariant superpotential 
of the Yukawa  sector respecting  $S_4$ and $Z_4$ symmetries is given as
\begin{align}
w_\text{$SU(5)$}^{(0)} &= y_1^u(T_1,T_2)\otimes (T_1,T_2)\otimes \chi _1\otimes H_5/\Lambda \nonumber \\
&\ + y_2^u(T_1,T_2)\otimes (T_1,T_2)\otimes (\chi _2,\chi _3)\otimes H_5/\Lambda + y_3^uT_3\otimes T_3\otimes H_5 \nonumber \\
&\ + M_1(N_e^c,N_\mu ^c)\otimes (N_e^c,N_\mu ^c) + M_2N_\tau ^c\otimes N_\tau ^c \nonumber \\
&\ + y^N(N_e^c,N_\mu ^c)\otimes (N_e^c,N_\mu ^c)\otimes (\chi _4,\chi _5) \nonumber \\
&\ + y_1^D(N_e^c,N_\mu ^c)\otimes (F_1,F_2,F_3)\otimes (\chi _6,\chi _7,\chi _8)\otimes H_5/\Lambda \nonumber \\
&\ + y_2^DN_\tau ^c\otimes (F_1,F_2,F_3)\otimes (\chi _6,\chi _7,\chi _8)\otimes H_5/\Lambda \nonumber \\
&\ + y_1(F_1,F_2,F_3)\otimes (T_1,T_2)\otimes (\chi _9,\chi _{10},\chi _{11})\otimes H_{\bar 5}/\Lambda  \nonumber \\
&\ + y_2(F_1,F_2,F_3)\otimes T_3\otimes (\chi _{12},\chi _{13},\chi _{14})\otimes H_{\bar 5}/\Lambda ,
\end{align}
 where  $M_1$ and $M_2$ are mass parameters
 for right-handed Majorana neutrinos,
and  Yukawa coupling constants $y_i^a$ and $y_i$ are complex in general.
By decomposing this superpotential into the quark sector and 
the lepton sector,
we can discuss mass matrices of quarks and leptons in  following sections.

%%%%%%%%%%%%%%%%%%%%%%%%%%%%%%%%%%%%%%%%%%%%%%%%%%%%%%%%%%%%%%%%%%%%%%%%
%%%%%%%%%%%%%%%%%%%%%%%%%%%%%%%%%%%%%%%%%%%%%%%%%%%%%%%%%%%%%%%%%%%%%%%%

\section{Lepton sector}
At first, we begin to discuss the  lepton sector of 
the superpotential $w_\text{SU(5)}^{(0)}$.
 Denoting Higgs doublets as $h_u$ and $h_d$,
 the superpotential of the Yukawa sector respecting 
the $S_4 \times Z_4$ symmetry  is given for charged leptons as
\begin{align}
w_l &= y_1\left [\frac{e^c}{\sqrt2}(l_\mu \chi _{10}-l_\tau \chi _{11})+\frac{\mu ^c}{\sqrt 6}(-2l_e \chi _{9}+l_\mu \chi _{10}+l_\tau \chi _{11})
\right ] h_d/\Lambda \nonumber \\
&\ + y_2\tau ^c ( l_e \chi _{12}+l_\mu \chi _{13}+l_\tau \chi _{14})h_d/\Lambda .
\end{align}
For  right-handed Majorana neutrinos,  the superpotential is given as
\begin{align}
w_N &= M_1(N_e^cN_e^c+N_\mu ^cN_\mu ^c) + M_2N_\tau ^cN_\tau ^c \nonumber \\
&\ +y^N\left[(N_e^cN_\mu ^c+N_\mu ^cN_e^c)\chi _4+(N_e^cN_e^c-N_\mu ^cN_\mu ^c)\chi _5\right ],
\end{align}
and for  neutrino Yukawa couplings,  the superpotential is
\begin{align}
w_D &= y_1^D\left [\frac{N_e^c}{\sqrt2}(l_\mu \chi _7-l_\tau \chi _8) +\frac{N_\mu ^c}{\sqrt 6}(-2l_e \chi _6 +l_\mu \chi _7+l_\tau \chi _8)\right ]
 h_u/\Lambda \nonumber \\
&\ +y_2^DN_\tau ^c(l_e\chi _6+l_\mu \chi _7+l_\tau \chi _{8})h_u/\Lambda .
\end{align}

 Higgs doublets $h_u,h_d$ and gauge singlet scalars $\chi _i$, 
are assumed to develop their vacuum expectation values (VEVs) as follows:
\begin{align}
&\langle h_u\rangle =v_u,
\quad
\langle h_d\rangle =v_d,
\quad
\langle (\chi _4,\chi _5)\rangle =(u_4,u_5),
\quad
\langle (\chi _6,\chi _7,\chi _8)\rangle =(u_6,u_7,u_8), \nonumber \\
&\langle (\chi _9,\chi _{10},\chi _{11})\rangle =(u_9,u_{10},u_{11}),
\quad 
\langle (\chi _{12},\chi _{13},\chi _{14})\rangle =(u_{12},u_{13},u_{14}),
\end{align}
which are supposed to be real.
Then, we obtain the mass matrix for charged leptons as
\begin{equation}
M_l = y_1v_d\begin{pmatrix}
                          0 & \alpha _{10}/\sqrt 2 & -\alpha _{11}/\sqrt 2 \\
                          -2\alpha _9/\sqrt 6 & \alpha _{10}/\sqrt 6 & \alpha _{11}/\sqrt 6   \\
                          0 & 0 & 0 \end{pmatrix}
+y_2v_d\begin{pmatrix}
               0 & 0 & 0 \\
               0 & 0 & 0 \\
               \alpha _{12} & \alpha _{13} & \alpha _{14} \\
          \end{pmatrix},
\label{charged}
\end{equation}
while the right-handed Majorana neutrino mass matrix is given as
\begin{equation}
M_N = \begin{pmatrix}
               M_1+y^N\alpha _5\Lambda & y^N\alpha _4\Lambda & 0 \\
               y^N\alpha _4\Lambda & M_1-y^N\alpha _5\Lambda & 0 \\
               0 & 0 & M_2
         \end{pmatrix},
\label{majorana}
\end{equation}
and the Dirac mass matrix of neutrinos is
\begin{equation}
M_D = y_1^Dv_u\begin{pmatrix}
                          0 & \alpha _7/\sqrt 2 & -\alpha _8/\sqrt 2 \\
                          -2\alpha _6/\sqrt 6 & \alpha _7/\sqrt 6 & \alpha _8/\sqrt 6   \\
                          0 & 0 & 0 \end{pmatrix}+y_2^Dv_u\begin{pmatrix}
                                                          0 & 0 & 0 \\
                                                          0 & 0 & 0 \\
                                        \alpha _6 & \alpha _7 & \alpha _8
                                                     \end{pmatrix},
\label{dirac}
\end{equation}
where we denote $\alpha_i \equiv  u_i/\Lambda$.

%%%%%%%%%%%%%%%%%%%%%%%%%%%%%%%%%%%%%%%%%%%%%%%%%%%
%%%%%%%%%%%%  Charged Lepton Mixing %%%%%%%%%%%%%%%
%%%%%%%%%%%%%%%%%%%%%%%%%%%%%%%%%%%%%%%%%%%%%%%%%%%

Let us discuss  lepton masses and mixing angles by considering mass matrices
in   Eqs.(\ref{charged}), (\ref{majorana}) and (\ref{dirac}).
In order to get the left-handed mixing of  charged leptons,
we investigate $M_l^\dagger M_l$.
If we can take   vacuum alignments 
$(u_{9}, u_{10},  u_{11})=(u_{9}, u_{10}, 0)$ and 
$(u_{12}, u_{13},  u_{14})=(0,0, u_{14})$, that is
 $\alpha_{11} = \alpha_{12} = \alpha_{13} = 0$,
we obtain 
\begin{equation}
M_l ^\dagger M_l = v_d^2
\begin{pmatrix}
\frac{2}{3}|y_1|^2\alpha _9^2 & -\frac{1}{3}|y_1|^2\alpha _9\alpha _{10} & 0 \\
-\frac{1}{3}|y_1|^2\alpha _9\alpha _{10} & \frac{2}{3}|y_1|^2\alpha _{10}^2 & 0 \\
0 & 0 & |y_2|^2\alpha _{14}^2
 \end{pmatrix},
\end{equation}
which gives
 $\theta^l_{13}=\theta^l_{23}=0$, where $\theta^l_{ij}$ denote
 left-handed mixing angles to diagonalize the charged lepton mass matrix.
Since the electron mass is tiny compared with the muon mass,
 we expect  $\alpha _9 \ll \alpha_{10}$ and then we get the mixing angle
$\theta_{12}^l$ as,
\begin{align}
\tan \theta^l_{12} &\approx -\frac{\alpha _9}{2\alpha _{10}},
\end{align}
 and  charged lepton masses, 
\begin{align}
m_e^2\approx \frac{1}{2}|y_1|^2\alpha _9^2v_d^2 \ ,
\ 
m_\mu ^2\approx 
\frac{2}{3}|y_1|^2\alpha _{10}^2v_d^2+\frac{1}{6}|y_1|^2\alpha _9^2v_d^2\approx \frac{2}{3}|y_1|^2\alpha _{10}^2v_d^2\ ,
\ 
m_\tau ^2=|y_2|^2\alpha _{14}^2v_d^2\ .
\label{chargemass}
\end{align}
Therefore, the mixing of $\theta^l_{12}$ is estimated as
\begin{align}
|\tan \theta^l_{12}|\approx \frac{1}{\sqrt 3}&\frac{m_e}{m_\mu}\approx 
2.8\times 10^{-3},
\label{leptonmix}
\end{align}
which is negligibly small.
Hereafter, we do not consider the mixing from the charged lepton sector.

%%%%%%%%%%%%%%%%%%%%%%%%%%%%%%%%%%%%%%%%%%%%%%%%%%%
%%%%%%%%%%%%  Majorana MR %%%%%%%%%%%%%%%%%%%%%%%%%
%%%%%%%%%%%%%%%%%%%%%%%%%%%%%%%%%%%%%%%%%%%%%%%%%%%

 Taking vacuum alignments $(u_{4}, u_{5})=(0, u_{5})$
and   $(u_{6}, u_{7}, u_{8})=(u_{6}, u_{6}, u_{6})$ in Eq.(\ref{majorana}).
By using the seesaw mechanism $M_\nu = M_D^TM_N^{-1}M_D$, 
the left-handed Majorana neutrino mass matrix is  written as
\begin{equation}
M_\nu = \begin{pmatrix}
                 a+\frac{2}{3}b & a-\frac{1}{3}b & a-\frac{1}{3}b \\
                 a-\frac{1}{3}b & a+\frac{1}{6}b+\frac{1}{2}c & a+\frac{1}{6}b-\frac{1}{2}c \\
                 a-\frac{1}{3}b & a+\frac{1}{6}b-\frac{1}{2}c & a+\frac{1}{6}b+\frac{1}{2}c
            \end{pmatrix},
\label{neutrino}
\end{equation}
where
\begin{equation}
a = \frac{(y_2^D\alpha _6v_u)^2}{M_2},\qquad 
b = \frac{(y_1^D\alpha _6v_u)^2}{M_1-y^N\alpha _5\Lambda },\qquad 
c = \frac{(y_1^D\alpha _6v_u)^2}{M_1+y^N\alpha _5\Lambda }.
\end{equation}
The neutrino mass matrix is decomposed as
\begin{equation}
M_\nu = \frac{b+c}{2}\begin{pmatrix}
                                  1 & 0 & 0 \\
                                  0 & 1 & 0 \\
                                  0 & 0 & 1
                             \end{pmatrix} + \frac{3a-b}{3}\begin{pmatrix}
                                                   1 & 1 & 1 \\
                                                   1 & 1 & 1 \\
                                                   1 & 1 & 1
                \end{pmatrix} + \frac{b-c}{2}\begin{pmatrix}
                                 1 & 0 & 0 \\
                                 0 & 0 & 1 \\
                                 0 & 1 & 0
                                            \end{pmatrix},
\end{equation}
which gives the tri-bimaximal mixing matrix 
$U_\text{tri-bi}$ and mass eigenvalues  as follows:
\begin{equation}
U_\text{tri-bi} = \begin{pmatrix}
               \frac{2}{\sqrt{6}} &  \frac{1}{\sqrt{3}} & 0 \\
     -\frac{1}{\sqrt{6}} & \frac{1}{\sqrt{3}} &  -\frac{1}{\sqrt{2}} \\
      -\frac{1}{\sqrt{6}} &  \frac{1}{\sqrt{3}} &   \frac{1}{\sqrt{2}}
         \end{pmatrix},
\qquad m_1 = b\ ,\qquad m_2 = 3a\ ,\qquad m_3 = c\ .
\end{equation}

%%%%%%%%%%%%%%%%%%%%%%%%%%%%%%%%%%%%%%%%%

\section{Quark sector}
In this section, we discuss quark sector of 
the superpotential $w_\text{SU(5)}^{(0)}$.
For up-type quarks, the superpotential of the Yukawa sector with
$S_4 \times Z_4$  is given as
\begin{align}
w_u &= y_1^u( u^cq_1+c^cq_2)\chi _1 h_u/\Lambda \nonumber \\
&\ +y_2^u\left[ ( u^cq_2+c^cq_1) \chi _2+(u^cq_1- c^cq_2)\chi _3\right ]
h_u/\Lambda 
 +y_3^u t^c q_3h_u.
\end{align}
For down-type quarks, we can write the superpotential as follows:
\begin{align}
w_d &= y_1\left [\frac{ 1}{\sqrt 2}( s^c\chi _{10} - b^c \chi _{11}) q_1+\frac{1}{\sqrt 6}(-2d^c \chi _9+s^c\chi _{10}+b^c\chi _{11})q_2 \right ]
h_d/\Lambda \nonumber \\
&\ + y_2( d^c\chi _{12}+ s^c \chi _{13}+  b^c\chi _{14})q_3 h_d/\Lambda .
\end{align}
We assume 
that scalar fields,  $\chi_i$, develop their VEVs as $u_i$.
%follows:
%\begin{align}
%&\langle \chi _1\rangle =u_1,
%\quad 
%\langle (\chi _2,\chi _3)\rangle =(u_2,u_3),
%\nonumber \\
%&\langle (\chi _9,\chi _{10},\chi _{11})\rangle =(u_9,u_{10},u_{11}),
%\quad
%\langle (\chi _{12},\chi _{13},\chi _{14})\rangle =(u_{12},u_{13},u_{14}).
%\end{align}
Then,  the mass matrix for up-type quarks is given as
\begin{equation}
M_u = v_u\begin{pmatrix}
                  y_1^u\alpha _1+y_2^u\alpha _3 & y_2^u\alpha _2 & 0 \\ 
                  y_2^u\alpha _2 & y_1^u \alpha _1-y_2^u\alpha _3 & 0 \\
                  0 & 0 & y_3^u
             \end{pmatrix},
\end{equation}
and  the down-type quark mass matrix is given as
\begin{equation}
M_d = y_1v_d\begin{pmatrix}
                      0 & -2\alpha _9/\sqrt 6 & 0 \\ 
                      \alpha _{10}/\sqrt 2 & \alpha _{10}/\sqrt 6  &  0 \\
                      -\alpha _{11}/\sqrt 2  & \alpha _{11}/\sqrt 6 & 0
                  \end{pmatrix} + y_2v_d\begin{pmatrix}
                                                     0 & 0 & \alpha _{12} \\ 
                                                     0 & 0 & \alpha _{13} \\
                                                     0 & 0 & \alpha _{14}
                                                \end{pmatrix}.
\end{equation}

Let us discuss %masses and 
mixing of the quark sector.
For up-type quarks, if we take 
\begin{equation}
\alpha _3 = 0,
\qquad 
y_1^u\alpha _1 = y_2^u\alpha _2,
\label{cond-up}
\end{equation}
which will be reexamined to get observed  CKM mixing  angles 
 in section 5.2, then we have
\begin{equation}
M_u = v_u\begin{pmatrix}
                 y_1^u\alpha _1 & y_1^u\alpha _1 & 0 \\
                 y_1^u\alpha _1 & y_1^u\alpha _1 & 0 \\
                 0 & 0 & y_3^u
             \end{pmatrix} ,
\end{equation}
which is diagonalized by the orthogonal matrix $U_u$
\begin{align}
U_u = \begin{pmatrix}
        \cos 45^\circ & \sin 45^\circ & 0 \\
        -\sin 45^\circ & \cos 45^\circ & 0 \\
        0 & 0 & 1
         \end{pmatrix}.
\end{align}
%The up-type quark masses   are given as 
%\begin{align}
%m_u=0,
%\qquad &m_c=2y_1^uv_u\alpha _1,
%\qquad m_t=y_3^uv_u.
%\end{align}

For down-type quarks, 
putting $\alpha _{11} = \alpha _{12} = \alpha _{13} = 0$,
which is the condition in  the charged lepton sector, we have
\begin{equation}
M_d^\dagger M_d = v_d^2\begin{pmatrix}
                                 \frac{1}{2}|y_1|^2\alpha _{10}^2 & 
\frac{1}{2\sqrt 3}|y_1|^2\alpha _{10}^2 & 0 \\
                                     \frac{1}{2\sqrt 3}|y_1|^2\alpha _{10}^2 & \frac{1}{6}|y_1|^2(4\alpha _9^2+\alpha _{10}^2) & 0 \\
                                     0 & 0 & |y_2|^2\alpha _{14}^2
                                 \end{pmatrix}.
\end{equation}
Then, the mass matrix is  diagonalized  by the orthogonal matrix $U_d$ as
\begin{align}
U_d = \begin{pmatrix}
            \cos 60^\circ & \sin 60^\circ & 0 \\
            -\sin 60^\circ & \cos 60^\circ & 0 \\
            0 & 0 & 1
         \end{pmatrix},
\label{Ud}
\end{align}
where  the small  $\alpha_9$ is neglected. 

Now, we get  the CKM matrix as follows:
\begin{equation}
V^{CKM} = U_u^\dagger U_d = \begin{pmatrix}
                                      \cos 15^\circ & \sin 15^\circ & 0 \\
                                      -\sin 15^\circ & \cos 15^\circ & 0 \\
                                           0 & 0 & 1
                                       \end{pmatrix}.
\end{equation}
Therefore, in our prototype model of $SU(5)$ GUT with the $S_4$ flavor symmetry,
 the quark sector has a non-vanishing mixing angle $15^\circ$
 only between the first and second generations
while the lepton flavor mixing is tri-bimaximal.
In order to get the non-vanishing but small mixing angles 
 $V_{cb}^{CKM}$ and $V_{ub}^{CKM}$, we  improve the prototype  model
 in the next section.

\section{Improved $S_4$ flavor model in $SU(5)$ GUT}
We improve the prototype model to get  the observed 
quark and lepton mass spectra  and the CKM mixing matrix.
We introduce
 the  $SU(5)$ $45$-dimensional Higgs $h_{45}$, which is required 
to get the difference between the charged lepton mass spectrum
 and the down-type quark mass spectrum. 
Moreover, we add   an $S_4$ doublet
 $(\chi _2^\prime ,\chi _3^\prime )$ and  an $S_4$ triplet
 $(\chi _9^\prime ,\chi _{10}^\prime ,\chi _{11}^\prime )$, 
which are  $SU(5)$ gauge singlet scalars.
These  assignments of $SU(5)$, $S_4$, and $Z_4$ are summarized Table 2.
Since the additional scalars do not contribute  to the neutrino sector,
 the result of the neutrino sector in the prototype model is not changed.
Therefore, we discuss only the  charged lepton sector and the quark sector
 in this section.
\begin{table}[h]
\begin{center}
\begin{tabular}{|c|c||cc|}
\hline
& $h_{45}$ & $(\chi _2^\prime ,\chi _3^\prime )$ & $(\chi _9^\prime ,\chi _{10}^\prime ,\chi _{11}^\prime )$ \\ \hline
$SU(5)$ & $45$ & $1$ & $1$ \\
$S_4$ & $1_1$ & $2$ & $3_1$ \\
$Z_4$ & $\omega^2$ & $\omega^3$ & $\omega ^2$ \\
\hline
\end{tabular}
\caption{Assignments of additional scalars in
$SU(5)$, $S_4$, and $Z_4$ representations.}
\end{center}
\end{table}

The superpotential of the Yukawa sector respecting the $SU(5)$, $S_4$ and $Z_4$ 
symmetries  is given as
\begin{equation}
w_\text{$SU(5)$} = w_\text{$SU(5)$}^{(0)}+w_\text{$SU(5)$}^{(1)},
\end{equation}
where we denote
\begin{align}
w_\text{$SU(5)$}^{(1)} &= y_4^u(T_1,T_2)\otimes T_3\otimes (\chi _2^\prime ,\chi _3^\prime )\otimes H_5/\Lambda \nonumber \\
&\ + y_1^\prime (F_1,F_2,F_3)\otimes (T_1,T_2)\otimes (\chi _9^\prime ,\chi _{10}^\prime ,\chi _{11}^\prime )\otimes h_{45}/\Lambda .
\end{align}

\subsection{Improved lepton sector}
Masses  and mixing angles of the charged lepton sector are 
  similar to those of  the prototype model in Eqs.(\ref{chargemass}) and
(\ref{leptonmix}). 
If we can take  the vacuum alignments 
$(u_{9}, u_{10},  u_{11})=(u_{9}, u_{10}, 0)$,
$(u_{9}^\prime, u_{10}^\prime,  
u_{11}^\prime)=(u_{9}^\prime, u_{10}^\prime, 0)$  and 
$(u_{12}, u_{13},  u_{14})=(0,0, u_{14})$, that is
 $\alpha_{11} = \alpha_{11}^\prime=\alpha_{12} =\alpha_{13} = 0$,
we obtain charged lepton mass matrix as follow:
\begin{equation}
M_l = v_d\begin{pmatrix}
            0 & (y_1\alpha _{10}-3\bar y_1\alpha _{10}^\prime )/\sqrt 2 & 0 \\
            -2(y_1\alpha _9-3\bar y_1\alpha _9^\prime )/\sqrt 6 & (y_1\alpha _{10}-3\bar y_1\alpha _{10}^\prime )/\sqrt 6 & 0 \\
            0 & 0 & y_2\alpha _{14}
         \end{pmatrix},
\end{equation}
where we replace  $y_1^\prime v_{45}$ with $\bar y_1v_d$. 
Masses and mixing angles of the charged leptons as follows:
\begin{eqnarray}
\hspace{-1cm}
&&m_e^2\approx\frac{1}{2}|y_1\alpha _9-3\bar y_1\alpha _9^\prime |^2v_d^2,
\quad 
m_\mu ^2\approx
\frac{2}{3}|y_1\alpha _{10}-3\bar y_1\alpha _{10}^\prime |^2v_d^2, 
\quad
m_\tau ^2\approx |y_2|^2\alpha _{14}^2v_d^2, \nonumber\\
&&
|\theta^l_{12}|=\left |-\frac{y_1\alpha _9-3\bar y_1\alpha _9^\prime }{2(y_1\alpha _{10}-3\bar y_1\alpha _{10}^\prime )}\right | 
\approx \frac{1}{\sqrt 3}\frac{m_e}{m_\mu }\approx 2.8\times 10^{-3},
\quad 
\theta^l_{23}=0,
\quad
\theta^l_{13}=0.
\end{eqnarray}
Thus, the charged lepton mass matrix is almost diagonal, 
and so the tri-bimaximal mixing of neutrino flavors
 is also  realized  in this improved model.

%%%%%%%%%%%%%%%%%%%%%%%%%%%%%%%%%%%%%%%%%%%
%%%%%%%%%% Improved Quark Sector %%%%%%%%%%
%%%%%%%%%%%%%%%%%%%%%%%%%%%%%%%%%%%%%%%%%%%

\subsection{Improved quark sector}
Let us discuss the quark sector of the superpotential $w_\text{SU(5)}$.
For up-type quarks, the mass matrix is given as
\begin{equation}
M_u = v_u\begin{pmatrix}
                  y_1^u\alpha _1+y_2^u\alpha _3 & y_2^u\alpha _2 & y_4^u\alpha _2^\prime \\ 
                  y_2^u\alpha _2 & y_1^u \alpha _1-y_2^u\alpha _3 & y_4^u\alpha _3^\prime \\
                  y_4^u\alpha _2^\prime & y_4^u\alpha _3^\prime & y_3^u
             \end{pmatrix},
\label{Mu}
\end{equation}
while the  down-type quark mass matrix is given as
\begin{align}
M_d &= y_1v_d\begin{pmatrix}
                      0 & -2\alpha _9/\sqrt 6 & 0 \\ 
                      \alpha _{10}/\sqrt 2 & \alpha _{10}/\sqrt 6  &  0 \\
                      -\alpha _{11}/\sqrt 2  & \alpha _{11}/\sqrt 6 & 0
                   \end{pmatrix} + y_2v_d\begin{pmatrix}
                                                     0 & 0 & \alpha _{12} \\ 
                                                     0 & 0 & \alpha _{13} \\
                                                     0 & 0 & \alpha _{14}
                                                 \end{pmatrix} \nonumber \\
&\ + y_1^\prime v_{45}\begin{pmatrix}
                                   0 & -2\alpha _9^\prime /\sqrt 6 & 0 \\ 
                                   \alpha _{10}^\prime /\sqrt 2 & \alpha _{10}^\prime /\sqrt 6  &  0 \\
                                  -\alpha _{11}^\prime /\sqrt 2  & \alpha _{11}^\prime /\sqrt 6 & 0
                               \end{pmatrix}.
\label{Md}
\end{align}

We consider the quark mixing.
The up-type quark mass matrix (\ref{Mu})
turns to the following one 
after rotating by $\theta_{12}^u=45^\circ$:
\begin{equation}
\hat M_u = v_u\begin{pmatrix}
                y_1^u\alpha _1-y_2^u\alpha _2 & y_2^u\alpha _3 & \frac{y_4^u}{\sqrt 2}(\alpha _2^\prime -\alpha _3^\prime ) \\
                y_2^u\alpha _3 & y_1^u\alpha _1+y_2^u\alpha _2 & \frac{y_4^u}{\sqrt 2}(\alpha _2^\prime +\alpha _3^\prime ) \\
                \frac{y_4^u}{\sqrt 2}(\alpha _2^\prime -\alpha _3^\prime ) & \frac{y_4^u}{\sqrt 2}(\alpha _2^\prime +\alpha _3^\prime ) & y_3^u
             \end{pmatrix}.
\end{equation}
In order to obtain the non-vanishing quark mixing of 
$V^{CKM}_{cb}$ and $V^{CKM}_{ub}$, we take
\begin{equation}
y_2^u\alpha _3\gg y_1^u\alpha _1 , \ y_2^u\alpha _2,
\qquad 
\alpha _2^\prime =\alpha _3^\prime ,
\end{equation}
which are realized by  vacuum alignments $u_{1}=0$
\footnote{One may consider to remove  
$\chi_1$, which is $S_4$ singlet, in our scheme.},
 $(u_{2}, u_{3})=(0, u_{3})$
and   $(u_{2}^\prime, u_{3}^\prime)=(u_{2}^\prime, u_{2}^\prime)$.
This situation of VEVs is completely different from 
that of the prototype model as seen in Eq.(\ref{cond-up}),
in which $V^{CKM}_{cb}$ and $V^{CKM}_{ub}$ vanish.
Then, we obtain  the so-called Fritzsch-type mass matrix \cite{Fritzsch}
\begin{equation}
\hat M_u \simeq v_u\begin{pmatrix}
                 0 & y_2^u\alpha _3 & 0 \\
                 y_2^u\alpha _3 & 0 & \sqrt 2y_4^u\alpha _2^\prime \\
                 0 & \sqrt 2y_4^u\alpha _2^\prime & y_3^u
             \end{pmatrix}.
\label{Fritzsch}
\end{equation}
As well known,
the complex phases in this $3\times 3$  matrix  can be removed 
by the phase matrix $P$  as  $P^\dagger \hat M_u P$;
\begin{equation}
  P =\begin{pmatrix}
     1 & 0& 0 \\ 0& e^{-i\rho}& 0\\ 0 & 0 & e^{-i\sigma}
             \end{pmatrix}.
\end{equation}
Therefore, up-type quark masses are
\begin{equation}
m_u=\left |\frac{y_3^u{y_2^u}^2\alpha _3^2}{2{y_4^u}^2{\alpha _2^\prime }^2}
\right | v_u ,
\qquad 
m_c=\left |-\frac{2{y_4^u}^2}{y_3^u}{\alpha _2^\prime }^2 \right |v_u
\qquad 
m_t=|y_3^u| v_u,
\end{equation}
and the mixing  matrix to diagonalize $\hat M_u$ in
 Eq.(\ref{Fritzsch}), 
$V_\text{F}~(M_u^\text{diagonal}=V_\text{F}^\dagger \hat M_uV_\text{F})$, is
\begin{equation}
V_\text{F}\approx \begin{pmatrix}
                   1 & \sqrt \frac{m_u}{m_c} & -\sqrt \frac{m_u}{m_t} \\
                  -\sqrt \frac{m_u}{m_c} & 1 & \sqrt \frac{m_c}{m_t} \\
                      \sqrt \frac{m_u}{m_t} & -\sqrt \frac{m_c}{m_t} & 1
                     \end{pmatrix}.
\end{equation}

The conditions from the lepton sector 
$\alpha_{11}=\alpha_{11}^\prime= \alpha_{12}=\alpha_{13}=0$ give
the down-type quark mass matrix:
\begin{equation}
M_d = v_d\begin{pmatrix}
            0 & -2(y_1\alpha _9+\bar y_1\alpha _9^\prime )/\sqrt 6 & 0 \\ 
            (y_1\alpha _{10}+\bar y_1\alpha _{10}^\prime )/\sqrt 2 & (y_1\alpha _{10}+\bar y_1\alpha _{10}^\prime )/\sqrt 6 & 0 \\
            0 & 0 & y_2\alpha _{14}
         \end{pmatrix},
\end{equation}
where we denote $\bar y_1v_d=y_1^\prime v_{45}$.
Then, down-type quark masses are  given as
\begin{align}
&m_d^2\approx \frac{1}{2}|y_1\alpha _9+\bar y_1\alpha _9^\prime |^2v_d^2\ ,
\quad  
m_s^2\approx
\frac{2}{3}|y_1\alpha _{10}+\bar y_1\alpha _{10}^\prime |^2v_d^2\ ,
\quad 
m_b^2\approx
 |y_2|^2\alpha _{14}^2v_d^2\ , 
\end{align}
and the mixing angle $\theta_{12}^d$ is  
$60^\circ+\delta\theta_{12}^d$, where
\begin{equation}
\delta \theta _{12}^d=-\frac{\sqrt 3|y_1\alpha _9+\bar y_1\alpha _9^\prime |^2}{4|y_1\alpha _{10}+\bar y_1\alpha _{10}^\prime |^2}=-\frac{m_d^2}{\sqrt 3m_s^2}\approx -1.5\times 10^{-3} .
\end{equation}
Therefore,  $\theta _{12}^d$ is almost $60^\circ$.

Let us  discuss  the CKM matrix.
The unitary matrices diagonalizing the up-type quark mass matrix and 
the down-type quark one, 
$U_u$ and $U_d$ are given, respectively,
\begin{eqnarray}
&&U_u = \begin{pmatrix}
                 \cos 45^\circ & \sin 45^\circ & 0 \\
                 -\sin 45^\circ & \cos 45^\circ & 0 \\
                 0 & 0 & 1
             \end{pmatrix}
\begin{pmatrix}
     1 & 0& 0 \\ 0& e^{-i\rho}& 0\\ 0 & 0 & e^{-i\sigma}
             \end{pmatrix}
\begin{pmatrix}  
                  1 & \sqrt \frac{m_u}{m_c} & -\sqrt \frac{m_u}{m_t} \\
                -\sqrt \frac{m_u}{m_c} & 1 & \sqrt \frac{m_c}{m_t} \\
                     \sqrt \frac{m_u}{m_t} & -\sqrt \frac{m_c}{m_t} & 1
                          \end{pmatrix} , 
\nonumber\\
&&U_d = \begin{pmatrix}
                         \cos 60^\circ & \sin 60^\circ & 0 \\
                        -\sin 60^\circ & \cos 60^\circ & 0 \\
                            0 & 0 & 1
                                                 \end{pmatrix}.
\end{eqnarray}
Since the CKM matrix is given as $U_u^\dagger U_d$, 
we obtain the relevant CKM mixing as
%elements are given as 
\begin{equation}
\left |V_{us}^\text{CKM}\right |=
\left | \sin 15^{\circ} - \cos 15^{\circ}\sqrt{\frac{m_u}{m_c}}e^{i\rho}\right |, \qquad
\left |V_{cb}^\text{CKM}\right |=\sqrt \frac{m_c}{m_t}, \qquad
\left |V_{ub}^\text{CKM}\right |=\sqrt \frac{m_u}{m_t}
.
\end{equation}
In the limit of neglecting the CP violating phase, $\rho=0$,
putting  typical values  at the GUT scale 
 $m_u=1.04\times 10^{-3}$GeV, $m_c=302\times 10^{-3}$ GeV, $m_t=129$GeV,
which are derived in Ref. \cite{Koide},  we predict
\begin{align}
\left |V_{us}^\text{CKM}\right |=0.202, \quad 
\left |V_{cb}^\text{CKM}\right |=0.048, \quad
\left |V_{ub}^\text{CKM}\right |=0.003
.
\end{align}
By adjusting the non-zero  phase $\rho=50^\circ$, %$-0.854758$
we can get the central value of the observed Cabbibo angle $0.226$.
Another phase $\sigma$ is still a free parameter.

\section{Summary}
We have presented  a flavor model with the  $S_4$ symmetry to unify
  quarks and leptons in the framework of the $SU(5)$ GUT.
 Three generations of $\overline 5$-plets in $SU(5)$ are assigned $3_1$ of 
$S_4$ while the  first and the second generations of 
$10$-plets  in  $SU(5)$  are assigned to be  $2$ of $S_4$,
and the third generation of $10$-plet is to be $1_1$ of $S_4$.
These  assignments of $S_4$ for $\overline 5$ and $10$ 
lead to the  %completely 
different structure 
of  quark and lepton mass matrices.
Right-handed neutrinos, which are $SU(5)$ gauge singlets, 
are also assigned $2$ for the first and second generations
and $1_1$ for  the third generation, respectively.
These  assignments are  essential to realize the tri-bimaximal mixing
of neutrino flavors.
 Vacuum alignments of scalars are also required to realize the tri-bimaximal 
mixing of neutrino flavors.
Our model predicts the quark  mixing  as well as the tri-bimaximal
mixing of leptons. Especially, the Cabbibo angle is
predicted to be $15^{\circ}$ in the limit of  the vacuum alignment.
We improve the model to predict  observed CKM mixing angles. 
% as well as the non-vanishing $U_{e3}$ of the neutrino flavor mixing.
The deviation from  $15^{\circ}$ in  $|V_{us}^\text{CKM}|$, the non-vanishing 
$|V_{cb}^\text{CKM}|$, and $|V_{ub}^\text{CKM}|$ are given by up-type quark masses.
%is given by
% $\sqrt{m_u/m_c}$, while the non-vanishing  $|V_{cb}^\text{CKM}|$ 
%and  $|V_{ub}^\text{CKM}|$ are given by  $\sqrt{m_c/m_t}$
% and   $\sqrt{m_u/m_t}$, respectively.
%The non-vanishing $U_{e3}$ of the neutrino flavor mixing
%is independent of these deviations.

%\section*{Acknowledgements}
%We would like to thank ...........

%\appendix
%\section{First Appendix} %Empty argument \section{} yields `Appendix'. 
%
%\section{Second Appendix}

\end{document}